\documentstyle[12pt,amssymbo]{article}
\newcommand{\vgl}[1]{eq.(\ref{#1})}
\begin{document}
\begin{titlepage}
\begin{flushright} Preprint-KUL-TF-92/10  \\march 1992
\\hepth@xxx/9203038
\end{flushright}
\vskip 2.cm
\begin{center}
{\large\bf Nonlocal regularisation and two-dimensional \\
induced actions }\\
\vskip 12.mm
{\bf F.~De Jonghe$^1$, R.~Siebelink $^2$ and W.~Troost $^3$}\\
\vskip 12.mm
Instituut voor Theoretische Fysica\\K.U.Leuven\\Celestijnenlaan 200
D\\B-3001 Leuven (Belgium)
\end{center}
\vskip 2.7cm
\begin{center}
{\bf Abstract}
\end{center}
\begin{quote}
We present an invariant regularisation scheme to compute two dimensional
induced gauge theory actions, that is local in Polyakov's variables, but
nonlocal in the original gauge potentials.  Our method sheds light on the
locality of this induced action, and leads to a straightforward proof that
the $\varepsilon$-anomaly in $W_3$-gravity is completely given by the one
loop term.
\vspace{4.cm}
\hrule width 16.cm  {\small
\noindent $^1$ Aspirant N.F.W.O.; E-mail: FGBDA16 at BLEKUL11 \\
\noindent $^2$ Aspirant N.F.W.O.; E-mail: FGBDA04 at BLEKUL11 \\
\noindent $^3$ Bevoegdverklaard Navorser
N.F.W.O.;E-mail: FGBDA19 at BLEKUL11 } \\
\normalsize
\end{quote}
\end{titlepage}
\newpage
In gauge theories, the regularisation scheme plays an essential
(though
often hidden) role in the study of anomalies, or induced gauge field
actions that arise when integrating out matter fields.  In particular, the
classical symmetries respected by the regularisation survive the
quantisation, whereas others are potentially anomalous.  One particular
scheme is the Pauli-Villars regularisation \cite{Pauli}.  There, if
one uses
covariant derivatives, the potential gauge symmetry breaking is reflected
in a non-invariant mass term. If one has a choice of mass
terms,
this may result \cite{Diaz} in a choice of which symmetries one keeps.
Furthermore,
these different choices can be related to each other by the addition of
(finite)
``counterterms''.  An explicit formula for this counterterm and explicit
examples are given in \cite{anombfv,Hatsuda}.

In this letter we present a somewhat unusual application of this choice of
regularisation scheme (mass term).  We will introduce, in specific
2-dimensional gauge theories, a class of mass terms related by a continuous
parameter $0 \leq \alpha \leq 1$.  The unusual feature of this mass term
will be, that it is non-local in terms of the ordinary gauge potential,
although it {\it is} local in terms of a different parametrisation (due to
Polyakov \cite{Polyakov}).  For $\alpha = 1$ the mass term will be
invariant, leading
straightforwardly to an invariant induced action.  For $\alpha = 0$, the
mass term will be non-invariant, but local.  The relation between the two
is given by the counterterm refered to above.  Thus this procedure can
actually
be turned into a way to (re)calculate the usual induced action. It gives
additional insight into why this action comes out to be {\it local} in
terms of the Polyakov variables. And finally, our method allows
for extension to more involved theories, like $W_n$-gravities.  We will use
it to show explicitly that the $\epsilon$-anomaly in $W_3$-gravity is
completely given by the one-loop term. Finally, we will give another
illustration of our method, computing the twodimensional gauge action
induced by fermions.

Let us, as a first example, consider $W_2$-gravity \cite{Polyakov}.  The
quantity
of interest is the action $\Gamma [h]$ for the gauge field $h$ induced by
matter fields $\varphi $ which are scalars:
\begin{eqnarray}
\exp - \Gamma [h] &=& \int {\cal D}\varphi \exp - S[\varphi
,h]\nonumber\\
S[\varphi ,h] &=& \frac{1}{2 \pi}\int
 d^2x(\partial\varphi.\bar\partial\varphi
-h\partial\varphi.\partial\varphi) .\label{defgamma}
\end{eqnarray}
The action $S$ is invariant under the combined transformation of $h$ and
$\varphi $.
\begin{eqnarray}
\delta _\epsilon \varphi &=&\epsilon \partial \varphi  \nonumber\\
\delta _\epsilon h       &=&\bar \partial \epsilon - h \partial \epsilon
+ \epsilon\partial h  . \label{epsilonsym}
\end{eqnarray}
To decide whether $\Gamma [h]$ is invariant, the formal argument on
\vgl{defgamma} is insufficient, as is well known.  We regularise the path
integral in \vgl{defgamma} in the Pauli-Villars spirit, by adding a
PV-field $\Phi $ transforming in the same way as $\varphi:
\delta_{\epsilon} \Phi = \epsilon \partial \Phi $.
There is an additional mass term proportional to $M^2$, which is taken to
infinity at the end.
\[S_{PV} [\Phi,h] = S[\varphi \rightarrow \Phi, h] + \mbox{mass term}.\]
Also, a $\Phi$ loop is given an extra minus sign \footnote{One may avoid
this sleight-of-hand by introducing two fermions and one boson with equal
kinetic operators. Since this point is irrelevant for the (mainly one loop)
considerations here, we will continue to use the customary language of
inserting a minus sign `by hand', although in fact all our considerations
are valid on a completely conventional Lagrangian level.}.

There are only one-loop diagrams, so the PV method simply regularises the
theory. Although it may appear that one doesn't have much freedom in
choosing a mass-term,
in fact one has. Any term of the form $\frac{1}{2} m^2(x)
\Phi^2(x)$ will do, as long as $m^2(x) \neq 0$. Let us now consider the
interesting possibilities for $W_2$ gravity. We can parametrise, after
Polyakov \cite{Polyakov},$$h= \bar\partial f/\partial f,$$ where  we
 implicitly
assume $\partial f \neq 0$. The reparameterisation invariance
of \vgl{epsilonsym}
corresponds to $\delta_\epsilon f= \varepsilon  \partial f$,
i.e.\ $\partial f$
is a density. As a consequence, taking $m^2(x) = M^2 \partial f$ actually
makes the mass term, and therefore also the regularisation, invariant~!
The induced action, calculated with this regularisation, is then
invariant for $\delta_{\epsilon} f$ (or $ \delta_{\epsilon} h$ in
\vgl{epsilonsym}), and is independent of $f$ (or $h$) as a result.

Now we discuss the relation between different regularisations. Two things
have to be specified: firstly the regularized Lagrangian (of PV type in our
case), i.e.\ the mass  term, and secondly one may wish to include a finite
counterterm. In our case, two regularisation schemes then correspond to
\begin{equation}
\exp - \Gamma ^{(i)}[h] = \int {\cal D} \varphi {\cal D} \Phi \exp \{-
S[\varphi,h] - S[\Phi,h]- S_M^{(i)} [\Phi]\} \exp - M^{(i)}[h]
\label{gammai} \end{equation}
where $i \in \{0,1\}$ labels the scheme, $S_M^{(i)}$ denotes the different
mass terms, and $M^{(i)}$ the finite counterterm, that depends on the gauge
field only. The condition that $\Gamma ^{(0)} = \Gamma ^{(1)}$ imposes a
relation between $M^{(0)}$ and $M^{(1)}$, that depends on the $S^{(i)}_M$.
If there exists an `interpolating' regularisation labeled by $\alpha $,
with mass term $S^{(\alpha )}_M = -\frac{1}{2} M^2 \Phi T(\alpha ) \Phi $,
then this relation is given by \cite{anombfv}
\begin{equation}
M^{(1)} [h] - M^{(0)}[h] = \frac{1}{2} \int^1_0 d \alpha\
\mbox{Tr}\left\{T^{-1}(\alpha) \frac{\partial T^{(\alpha )}}{\partial \alpha}
\exp \frac{{\cal R}(\alpha )}{M^2}\right\}
\label{emeen}
\end{equation}
where ${\cal R}$ is a Fujikawa-style regulator, which is unambiguously
fixed
by the PV Lagrangian ${\cal L}_{P.V.} [\Phi] = \frac{1}{2}  \Phi (T {\cal
R}) \Phi
- \frac{1}{2}  M^2 \Phi T \Phi$, and where the trace is over functional
space.
This formula is given in \cite{anombfv} in a somewhat different context, and
is foreshadowed in \cite{bouwknegtvn}. In all of the above, a limit $M^2
\rightarrow \infty$, possibly after a  renormalisation, is understood.

Let us apply it to our case, taking $T(\alpha ) = (\partial f)^\alpha $.
For $\alpha  = 1$, as argued above, there is no anomaly, and therefore
\[\Gamma [h] = M^{(1)} [h]\,.\]
For $ \alpha  = 0$, the regularisation is non-invariant (the anomaly). The
result for $\Gamma[h]$ will be the same in both regularisations if we
satisfy \vgl{emeen}. Thus
\begin{equation}
\Gamma [h] = M^{(0)}[h] + \frac{1}{2} \int^1_0
d\alpha\mbox{Tr}\left\{T^{-1}\frac{\partial T}{\partial \alpha } \exp
\frac{{\cal R}}{M^2}\right\}\,.
\label{action}
\end{equation}
The regularisation with $\alpha =0$, which is local in $h$, corresponds
closest to the usual prepossession that any `fundamental' (i.e.\
non-induced) action is local. Also, $M^{(0)}$ is then to be considered as a
possible additional action for $h$, but not part of an action induced by
the matterfields $\varphi$. Therefore, we set $M^{(0)} = 0$. Then
\vgl{action}
gives an explicit closed formula for the induced action, which we will
evaluate forthwith. From the action for $ \Phi$ one reads off
\begin{eqnarray*}
T(\alpha ) &=& \frac{1}{2\pi} (\partial  f)^\alpha \\
T{\cal R} &=& - \frac{1}{\pi} (\partial  \bar\partial  - \partial  h
\partial )\\
{\cal R}(\alpha ) &=& - 2(\partial f)^{-\alpha } (\partial \bar\partial
- \partial  h\partial )\\
T^{-1} \frac{\partial T}{\partial \alpha } &=& \log (\partial f) \quad .
\end{eqnarray*}
The functional trace can be evaluated by a variety of methods, but the
easiest is to look up the result in \cite{Gilkey}. One can rewrite the
differential operator ${\cal R}$ as a covariant Laplacian with $g^{\mu\nu}
=\left(\begin{array}{cc}
2h &  -1\\
-1 & 0
\end{array} \right) (\partial f)^{-\alpha }$. The result of the trace is
then
\begin{eqnarray*}
\lim_{M\rightarrow \infty} Tr \{\log(\partial  f) \exp \frac{{\cal
R}[g^{\mu\nu}]}{M^2}\} = \lim_{M\rightarrow \infty} \int d(vol)
\frac{1}{4\pi}\log(\partial f) \{M^2 - \frac{1}{6}
R[g^{\mu\nu} ]\}
\end{eqnarray*}
where $d(vol)$ and $ R[g^{\mu\nu}]$ are the invariant volume element and
the Riemann scalar corresponding to the metric $g^{\mu\nu}$. Observe
that
\begin{equation}
\sqrt{g} R[g^{\mu\nu}] = R[\tilde{g}^{\mu\nu}] + \frac{1}{2} \partial _\mu
\{\sqrt{g} g^{\mu\nu} \partial _\nu \log g\}
\label{er}
\end{equation}
where $\tilde{g}^{\mu\nu} = \sqrt{g} g^{\mu\nu} = \left(\begin{array}{cc}
2h &  -1\\
-1 & 0
\end{array} \right)$, and that $g^{\mu\nu}$ has a
curvature scalar equal
\linebreak
to $ 2 \partial ^2 h$.
The second term in \vgl{er} equals $-2\alpha  \partial ^2 h$. The
induced action of \vgl{action} then becomes
\begin{eqnarray*}
\Gamma [h] = \lim_{M\rightarrow \infty} \frac{1}{2}  \int^1_0 d\alpha
\frac{1}{4\pi} \int d^2x \log(\partial f)   2(\partial
f)^\alpha  [M^2 -
\frac{1}{6} (\partial f)^{-\alpha } 2(1-\alpha ) \partial ^2 h].
\end{eqnarray*}
The necessary renormalisation is performed by dropping the $M^2$ terms,
(more precisely, by considering in the traditional way several PV fields
with $\Sigma c_i = 1$ and $\Sigma c_i M^2_i = 0$.~\footnote{See footnote
1, and choose the masses of the PV boson and fermions appropriately.}), or
 alternatively by including a normalisation factor $[\int{\cal D}  \Phi \exp
 - S_M^{(i)}[\Phi]]^{-1}$ in \vgl{gammai} from the start. The final result
 for the induced action is
 then the well known \[\Gamma [h] = - \frac{1}{24\pi} \int d^2x (\partial^2
h) \log(\partial f) = - \frac{1}{24\pi} \int d^2x (\partial^2 h)
\frac{1}{\Box} (\partial^2 h).\]
Let us recapitulate how we obtained it: we started from the observation
that
different regularisations will give the same result for $\Gamma $ if one
adds counterterms satisfying \vgl{emeen}. Then we presented a scheme that
is
invariant, albeit non local in $h$. The induced part of $\Gamma $ vanishes
in this scheme, so that $\Gamma $ equals the corresponding finite
counterterm. It is then computed from \vgl{emeen}.

Although
locality in $h$ is not present, the whole treatment {\it is local in f}.
The fact that the final induced action is local in $f$ is quite mysterious
from alternative ways to compute $\Gamma $ (like Feynman diagrams for
instance).

Our technique offers a way to understand why, in more involved
2-dimensional gravities \cite{generalWn}, the $\epsilon $-anomaly is not
changed
by the presence of the couplings with higher spin fields~\cite{alexetal}.
First, these extra couplings are $\epsilon $-invariant by themselves. Then,
with the
help of the {\it covariant} derivative (we mimic here the two-step
procedure followed in non-abelian
gauge theories in \cite{FandS}), one can construct higher
derivative terms, that are local in $h$, not just in $f$.
 Being covariant explicitly, this does not introduce anomalies. We
 will check that, as is the case in nonabelian gauge theories, this
 regularises
all diagrams except 1-loop. Then the Pauli-Villars method for 1 loop
finishes off the divergences. This is where Polyakov's $f$ comes in. The
anomaly then arises in the transition from the invariant nonlocal
regularisation to the noninvariant local one, \vgl{action}, exactly
as in $W_2$.
We now prove these assertions.

We give as an explicit example the $W_3$ case.
Consider the classical action:
\begin{equation}
S[\phi,h,B] = \frac{1}{\pi} \int d^2x \{- \frac{1}{2} \bar{\partial }
\phi^i \partial \phi^i
+ \frac{1}{2} \partial  \phi^i h \partial \phi^i
+ \frac{1}{3} B d^{ijk}\partial \phi^i \partial \phi^j \partial \phi^k\}.
\label{classaction}
\end{equation}
This action is invariant under the $\epsilon $-symmetry, \vgl{epsilonsym} and
\begin{eqnarray*}
\delta _\epsilon  B = \epsilon  \partial  B - 2 B \partial  \epsilon .
\end{eqnarray*}
Remark that the $B$-interaction term is invariant by itself.
The regularising term to be added to \vgl{classaction} is
\begin{eqnarray}
S_{reg} &=&  - \frac{1}{\pi} \int d^2x \frac{2}{\Lambda ^2} \partial
\phi^i \bar{\bigtriangledown}^3 \phi^i\ ,\label{regulterm}\\
\bar{\bigtriangledown} &=& \bar{\partial } - h \partial.  \nonumber
\end{eqnarray}
This term is clearly $\epsilon $-invariant. (For an explicit treatment of the
 $\epsilon$-covariance, see for example \cite{alexetal}.)

The addition of \vgl{regulterm} results     in a modification (cut-off) of
the propagator, and in extra vertices. The $\varphi$-propagator becomes

\begin{eqnarray*}
\frac{1}{\bar{k} k} \rightarrow \frac{1}{\bar{k} k} \left(\frac{\Lambda
^2}{\Lambda ^2 - \bar{k}^2}\right)
\end{eqnarray*}
where we use a complex notation for the momentum vectors,
 $\vec{k} . \vec{x} = (\bar{k}z + k\bar{z})/2$. The new vertices all have
two $\varphi$-lines and a number of $h$-lines, and momentum factors.
With $L$ the number of loops, $i_\varphi$ the number of internal
 $\varphi$-lines, $n_H$ the number of $h$-vertices, $n_B$ the number of
 $B$-lines, and  $n_\varphi$ the number of external
 $\varphi$-lines we compute the superficial degree of divergence $D$ and
find the following relations:
 \begin{eqnarray*}
L &=&  1+ (n_B-n_\varphi)/2\\
2i_\varphi &=& 3 n_B + 2n_H -n_\varphi\\
D &\leq& 2-2 n_B .
\end{eqnarray*}
Clearly, the divergent diagrams have $n_B \leq 1$, and then the number of
 loops is one or zero. This proves that the addition of the covariant
higher derivative term \vgl{regulterm} has eliminated all divergences
except the single loops. Now duplicate the action for the PV-fields $\Phi
^i$. This renders the one-loop diagrams finite too. As shown above, one
can take an invariant mass term (nonlocal in $h$). The transition to the
non-invariant local mass term is made in the same way as before,
through \vgl{emeen}. This finishes the argument.

As a second example, we now apply our scheme to the calculation of the
induced action for a gauge field coupled to $n$ chiral fermions
\cite{Polyakov-Wiegman}. The action is
\[S[\psi,A] = \frac{1}{2\pi } \int d^2x \, \psi^t(\bar\partial-A)\psi \]
with $\psi$ a column of fermionic degrees of freedom, and $ A = A^a t_a $
with the $t_a$ the generators of a Lie algebra in a representation $R$.
This representation must be such
that $R\bigwedge R$ contains the adjoint representation.
The matter fields transform as
\begin{eqnarray*}
\delta  \psi &=&  \eta \psi\\
\delta  \psi^t &=& -\psi^t\eta
\end{eqnarray*}
with $\eta=\eta^a t_a$, and $\eta^a$ the transformationparameters.
If we take
\begin{eqnarray*}
\delta A &=& \bar\partial\eta + [\eta,A]
\end{eqnarray*}
the action remains invariant.

The path-integral is regularised by introducing PV-fields as follows:
\begin{eqnarray*}
S_{\mbox{\footnotesize Reg}} &=& S[\psi,A] + S[\xi_1,A] - \frac{1}{2\pi }
\int d^2x \, \xi_2^t \partial \xi_2
 + S_{\mbox{\footnotesize mass}}[\xi_1,\xi_2]
\end{eqnarray*}
Notice that, in order to construct a mass term, we have introduced another
PV-field of opposite chirality. This PV field does not transform under the
gauge transformation~\footnote{The remarks of footnote 1 are applicable
also
here}. The standard mass term, eventually leading to the anomaly, or the
WZWN induced action, is non-invariant:
\[S^{(0)}_{mass} [\xi_1,\xi_2] = - \frac{M}{4\pi} \int d^2x\,(  \xi_1^t
\xi_2 - \xi_2^t \xi_1). \]
In this case also, one can construct an invariant mass term. If one
reparametrises \cite{Polyakov-Wiegman}
$A = \bar\partial g.g^{-1}$ , and $\delta g=\eta g$, then
\[S_{mass}^{(1)} [\xi_1,\xi_2] = - \frac{M}{4\pi} \int d^2x \, ( \xi_1^t g
\xi_2 - \xi_2^t g^{-1} \xi_1)\]
is invariant, thus leading to a zero induced action. The reasoning leading
to \vgl{action} again applies, mutatis mutandis. The interpolating
regularisation is now given by
\begin{equation}
S_{mass}^{(\alpha )} [\xi_1,\xi_2] = - \frac{M}{4\pi} \int d^2x \, ( \xi_1^t
g^\alpha \xi_2 - \xi_2^t g^{-\alpha } \xi_1) \label{alfamassa}
\end{equation}
where we have assumed $g$ to be  connected to the identity, so that
$g^\alpha $ makes sense. The matrices $T(\alpha )$ and the matrix
differential operator ${\cal R}$ are again unambiguously fixed by the PV
Lagrangian. In the present case (see \cite{anombfv,Hatsuda}for details),
writing
\[S[\xi_1,A] + S[\xi_2] + S_{mass}^{(\alpha)} [\xi_1,\xi_2] =
\int d^2x \,\frac{1}{2}  (\xi_1^t \:, \xi_2^t )  T {\cal O}
\left(\begin{array}{c} \xi_1 \\ \xi_2 \end{array} \right)
- \frac{1}{2} M  (\xi_1^t \:, \xi_2^t )  T
\left(\begin{array}{c} \xi_1 \\ \xi_2 \end{array} \right) \]
we have that ${\cal R} = {\cal O}^2$ is the Fujikawa-style regulator.
More in detail
\begin{eqnarray*}
T(\alpha ) &=& \frac{1}{2\pi}
\left(\begin{array}{cc}
0 & g^{\alpha}\\
-g^{-\alpha } & 0
\end{array} \right)
\end{eqnarray*}
and putting $g = e^{\Theta} = e^{\theta^a t_a}$ we find that
\begin{eqnarray*}
T^{-1} (\alpha ) \frac{\partial }{\partial \alpha } T(\alpha
) &=& \left(\begin{array}{cc}
-\Theta & 0\\
0 & \Theta
\end{array} \right) \\
\mbox{and}\: \cal R &=&
\left(\begin{array}{cc}
4g^{\alpha} \partial g^{-\alpha} (\bar\partial-A) & 0 \\
0 & 4g^{-\alpha} (\bar\partial-A) g^{\alpha} \partial
\end{array} \right) .
\end{eqnarray*}
The induced action \vgl{action}, now becomes
\begin{eqnarray}
\Gamma [h] = \lim_{M^2\rightarrow \infty} - \frac{1}{2} \int^1_0 d \alpha
Tr \left( \Theta e^{4g^{\alpha} \partial g^{-\alpha} (\bar\partial - A)
/M^2} - \Theta e^{4g^{-\alpha} (\bar\partial-A) g^{\alpha} \partial /M^2}
\right) . \label{voorgilkey}
\end{eqnarray}
where Tr stands for taking the trace over function space and over matrices.
One can proceed by direct computation, but it is convenient at this point
to use cyclicity of the trace to rewrite the second term of \vgl{voorgilkey}
as
\begin{eqnarray}
Tr \left( - \Theta e^{4 (\bar\partial-A) g^{\alpha} \partial
g^{-\alpha}/ M^2} \right) .\nonumber
\end{eqnarray}
Using the covariant derivatives
\begin{eqnarray*}
\bigtriangledown &=& \partial + B \equiv \partial +g^\alpha
(\partial g^{-\alpha})\\
\bar{\bigtriangledown} &=& \bar{\partial} + \bar{B} \equiv \bar{\partial} -A
\end{eqnarray*}
one obtains
\begin{eqnarray*}
\Gamma [h] = \lim_{M^2\rightarrow \infty} - \frac{1}{2} \int^1_0 d \alpha
Tr \left(\Theta e^{4 \bigtriangledown\bar{\bigtriangledown} /M^2}
 - \Theta e^{4 \bar{\bigtriangledown}\bigtriangledown/M^2}
\right) .
\end{eqnarray*}

Again, one can save oneself a (non-negligible) amount of work by realising
that this type of functional trace can be treated with the help of
\cite{Gilkey}. Essentially
one rewrites the regulator in such a way that the second derivatives appear
covariantly and symmetrically, while the first derivatives are absorbed into
the
connection on the fibre.  If, as in our case, the space metric is flat,
the final result
becomes very simple.\\ One rewrites $4 \bigtriangledown\bar{\bigtriangledown}
= 2\{\bigtriangledown ,\bar{\bigtriangledown}\} -E_G$ and uses
  Theorem 4.2, (2)in \cite{Gilkey} to obtain
\begin{equation}
\Gamma [h] = - \frac{1}{2} \int_0^1 d\alpha \int d^2x
\mbox{tr}\left(\frac{\Theta }{4\pi } \left[ M^2 + E_G \right] -
\frac{\Theta }{4\pi } \left[ M^2 - E_G \right]\right)\,.
\label{nagilkey}\nonumber
\end{equation}
For our case, we obtain
\[E_G = 2(\overline{\partial }B - \partial \overline{B}
+ [\overline{B},B])\]
which is the curvature of the nonabelian $B,\overline{B}$ field. Finally:
\begin{eqnarray}
\Gamma [h] &=& -\frac{1}{2\pi}\int d^2x \int_0^1 d\alpha  \mbox{tr}
\Theta \left[\overline{\partial }(g^\alpha \partial g^{-\alpha }) +
\partial (\overline{\partial }g\cdot g^{-1})\right.\nonumber\\
& &\left. - \overline{\partial }g\cdot g^{-1}\cdot g^\alpha
\partial g^{-\alpha }
+ g^\alpha \partial g^{-\alpha }\cdot
\overline{\partial }g\cdot g^{-1}
\right]      .   \nonumber
\label{eerste resultaat}
\end{eqnarray}
This expression can be rewritten in a multitude of different ways.  A short
road to identify the known result \cite{Polyakov-Wiegman} as the
WZWN action \cite{WZWN} is as follows.  First, by
using the formula for the derivative of the exponential of an operator (see
for example \cite{Fried}) one can write, using $g = e^\Theta $,
\begin{eqnarray} \mbox{tr}(\partial g^{-1}\overline{\partial }g) &=&
-\mbox{tr}\left(\int_0^1 d\alpha\ g^{-1+\alpha }\cdot \partial
\Theta \cdot g^{-\alpha }\cdot \overline{\partial }g\right) = -\mbox{tr}\
\int_0^1 d\alpha \ g^\alpha \partial \Theta g^{-\alpha }\overline{\partial
} gg^{-1}\nonumber\\
&=& \mbox{tr}\int_0^1 d\alpha (-\partial (g^\alpha \Theta g^{-\alpha
})\overline{\partial} gg^{-1} + \partial g^\alpha \cdot \Theta \cdot
g^{-\alpha }\overline{\partial }gg^{-1}\nonumber\\
&&\hspace*{3cm}+ g^\alpha \Theta \partial g^{-\alpha }\cdot
\overline{\partial }g\cdot g^{-1})
\label{tussenresultaat 1}
\end{eqnarray}
in which one recognises the last three terms in \vgl{eerste resultaat}
(freely using partial integration, dropping the boundary terms, in the
2-dimensional space).  The second trick is to use $\partial _\alpha
g^\alpha \cdot g^{-\alpha } = \Theta $ in
\begin{eqnarray}
\mbox{tr}(\partial g^{-1}\cdot \overline{\partial }g) &=& \mbox{tr}
\int_0^1 d\alpha \ \partial _\alpha (\partial g^\alpha \overline{\partial
}g^{-\alpha }) = \mbox{tr}\int_0^1 d\alpha (\partial (\Theta g^\alpha
)\overline{\partial }g^{-\alpha } - \partial g^\alpha \overline{\partial}
(\Theta g^{-\alpha }))\nonumber\\
&=& \mbox{tr}\int_0^1 d\alpha \ \Theta (-\partial (g^\alpha
\overline{\partial }g^{-\alpha }) + \overline{\partial }(\partial g^\alpha
\cdot g^{-\alpha }))
\label{tussenresultaat 2}
\end{eqnarray}
to rewrite the first term of \vgl{eerste resultaat} as
\[\mbox{tr}\int_0^1 d\alpha (-\partial _\alpha g^\alpha \cdot g^{-\alpha
}[\overline{\partial }g^\alpha \cdot g^{-\alpha },\partial g^\alpha \cdot
g^{-\alpha }] + \Theta \partial (g^\alpha \overline{\partial }g^{-\alpha
}))\,.\]
By combining this with \vgl{tussenresultaat 2}, where one also finds back
the first term of \vgl{eerste resultaat}, to eliminate the left-over
$\Theta $-terms, one obtains, using also \vgl{tussenresultaat 1}
\begin{equation}
\Gamma [h] = - \frac{1}{4\pi }\mbox{tr}\left\{\int_0^1 d\alpha \ \partial
_\alpha g^\alpha [\partial g^\alpha \cdot g^{-\alpha },\overline{\partial
}g^\alpha \cdot g^{-\alpha }] + \ \partial
g^{-1}\overline{\partial }g\right\} \nonumber
\label{WZWN active}
\end{equation}
which is proportional to the celebrated WZWN action, the proportionality
constant depending on the representation. Note how the parameter $\alpha
$, which was introduced as an interpolating parameter between different
regularisations, has now turned into the `extra' dimension which is
conveniently introduced to write down the topological term.  The
3-dimensional volume of which the 2-dimensional $x$-space is the boundary
can be viewed as having $\alpha $ as the radial coordinate.  The
`extension' of $g(x)$ to the 3-dimensional volume is here simply
$g(x,\alpha ) = [g(x)]^\alpha $.

Again, as in the 2-dimensional induced gravity example, the result
\vgl{WZWN active} is well known.  Our derivation sheds new light on one
of the most surprising aspects of that result, viz.\ its locality in $g$.
This hinges crucially on three ingredients.  The first is the
reparametrisation $A = \overline{\partial }g\cdot g^{-1}$ (cf.\ $h =
\overline{\partial }f/\partial f)$.  The second is that, given this
parametrisation, it becomes possible to write down an {\it invariant
regularisation} (mass term) which is local in $g$ (resp. $f$).  And
finally, one makes use of
the fact that the functional trace calculations arising from \vgl{action}
always give local expressions \cite{Seeley}.  Needless to say, once the
locality in $g$ is established, the induced action is almost determined.

Thus, we have shown, at least in two dimensions, the use of (mildly) non-
local but invariant regularisations can be turned into a tool to compute
the anomalies associated to strictly local regularisations.  This can be
exploited for more complicated theories like $W_n$ gravities, to
regularise invariantly in a two-step process.   Namely, first we regularised
all $n
\geq 2$-loop
diagrams using higher derivatives, covariantly.  The remaining 1-loop
divergence is regularised
with the Pauli-Villars method.  This second step is taken in a way that is
non-local in the original variables, so that a
counterterm as in \vgl{action} is needed to produce the result of a local
regularisation.  This counterterm is then the only source of anomalies.
We showed explicitly on the example of $W_3$ how this proofs that the
one-loop $\epsilon $-anomaly is an all-loop result.
In default of a covariant calculus for the $\lambda $-symmetries, it is not
clear whether the above ideas can be adopted to the full symmetry algebra
of $W_n$.
\section*{Acknowledgments}
It is a pleasure to thank A. Sevrin for discussions on
two-dimensional
induced actions, A. Slavnov for a discussion on regularisation, and
S. Vandoren  and A. Van Proeyen for encouragement.

\end{document}